\documentclass[
    ,final            
  ]
  {aipproc}

\layoutstyle{6x9}

\begin{document}

\title{Planets around Giant Stars}


\author{A. Quirrenbach}{
  address={ZAH, Landessternwarte, K\"onigstuhl 12, 69117 Heidelberg}
}

\author{S. Reffert}{
  address={ZAH, Landessternwarte, K\"onigstuhl 12, 69117 Heidelberg}
}

\author{C. Bergmann}{
  address={ZAH, Landessternwarte, K\"onigstuhl 12, 69117 Heidelberg}
  }

\begin{abstract}
We present results from a radial-velocity survey of 373 giant stars at Lick Observatory, which started in 1999. The previously announced planets $\iota$\,Dra\,b and Pollux\,b are confirmed by continued monitoring. The frequency of detected planetary companions appears to increase with metallicity. The star $\nu$\,Oph is orbited by two brown dwarf companions with masses of 22.3\,M$_{\rm jup}$ and 24.5\,M$_{\rm jup}$ in orbits with a period ratio close to 6:1. It is likely that the two companions to $\nu$\,Oph formed in a disk around the star.
\end{abstract}

\maketitle


\section{Introduction}

Giant stars are interesting targets for planet searches for a number of reasons. First, intermediate-mass ($\approx 1.5 \ldots 5$\,M$_\odot$) stars are not good objects for precise radial-velocity surveys during their life as early F and A stars on the main sequence, but later they evolve into G or K giants with many sharp absorption lines, amenable to very accurate Doppler measurements. Second, the expansion of the host star during the ascent of the giant branch may alter the architecture of the planetary system. Short-period planets may get swallowed completely, and the orbits of somewhat more distant planets may get altered by tidal interactions. Third, as giants are intrinsically bright, one may hope to use them as probes for planetary systems in more distant stellar populations.

For each of these objectives, one needs to compare the number and properties of the detected planets with benchmark statistics obtained from the current large surveys targeting mainly late F, G, and K main-sequence stars. This task is complicated as all of the above effects (parent star mass, evolution, population) can enter simultaneously; selection effects due to specific sample selection criteria must therefore be taken into account properly. In addition, it is much more difficult to determine mass and evolutionary state for field giants than for main sequence stars. In spite of these complications, the abundance of heavy ($m > 5$\,M$_{\rm jup}$) companions found in giant star surveys indicates that these objects are indeed more frequent around intermediate-mass than around Solar-mass stars.

\section{The Lick K Giant Survey}

The Lick radial-velocity survey of giant stars was initiated in 1999 with the goal of finding out whether these stars are sufficiently stable to be used as reference stars for very precise astrometric measurements (Frink et al.\ 2001). The 179~stars in our original sample were selected from the Hipparcos catalog; additional information from the Tycho-2 catalog was used in the selection process as well. All are bona fide K~giant stars (luminosity class III), and brighter than $6^{th}$ magnitude, with the exception of 12~stars which have luminosity class I--II. Initially, 86~stars were selected which showed no signs of variability or duplicity (see Frink et al.\ 2001 for the detailed criteria), and these stars were observed beginning in June~1999. In June~2000, 96~stars were added to the sample with less stringent criteria. For example, the criteria against long-term proper motion differences and photometric variability were relaxed. Three stars were removed from the sample when it was discovered that they were visual binaries, leaving a total of 179~stars. In 2004, the survey sample was augmented by another 194 G and K giants. These stars are on average bluer and more massive than the stars in the original sample.

We have been using the 0.6\,m Coud\'e Auxiliary Telescope (CAT) in conjunction with the Hamilton Echelle Spectrograph and an iodine cell in the beam. Our observing and data reduction techniques follow the procedures described by Butler et al.\ (1996). Exposure times vary between 20\,s and 1800\,s, depending on the brightness of the star. We obtain typical uncertainties of individual measurements in the range $5 \ldots 8$\,m/s, limited by the signal-to-noise ratio in our exposures. It would be possible to obtain smaller error bars by exposing each star longer, but this would be of limited value since our errors are already smaller than the intrinsic photospheric ``noise'' due to stellar oscillations even in the quietest stars in our sample (Hekker et al.\ 2006, see also Fig.~\ref{Hekker}).

\begin{figure}
  \includegraphics[width=.8\textwidth]{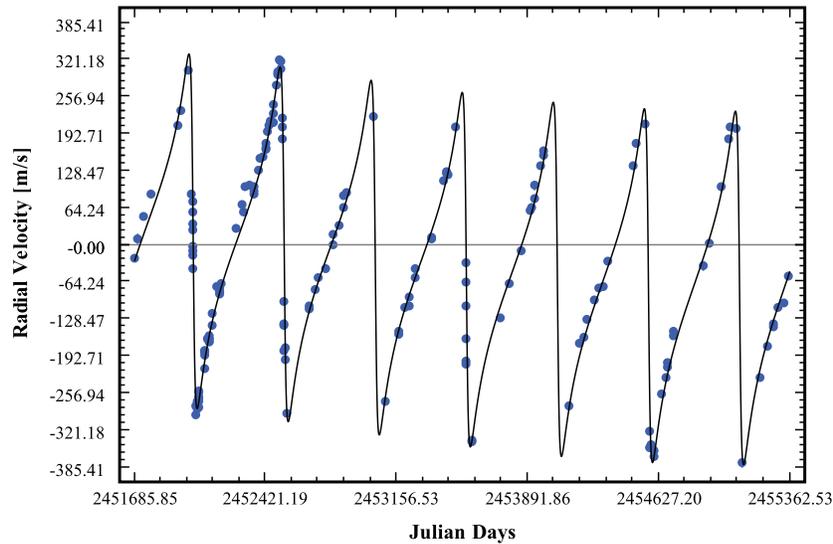}
  \caption{\label{iotdra} Radial-velocity data for $\iota$\,Dra (Frink et al.\ 2002, Zechmeister et al.\ 2008, and new measurements). The fit is a Keplerian ($p = 511$\,d, $e = 0.71$, $m \sin i = 10.1$\,M$_{\rm jup}$) plus a linear trend.}
\end{figure}

The Lick survey yielded the first confirmed planet around a giant star, namely $\iota$\,Dra (Frink et al.\ 2002, see also Fig.~\ref{iotdra}). In this case the Keplerian nature of the radial-velocity signal could be established easily because of the large eccentricity, and the substellar nature of $\iota$\,Dra\,b could be confirmed by the absence of any astrometric signature in the HIPPARCOS intermediate data. Continued monitoring has fully confirmed the reality of the planet, and in addition revealed a linear radial-velocity trend, which is likely due to a third body in the system.

\begin{figure}
  \includegraphics[width=.85\textwidth]{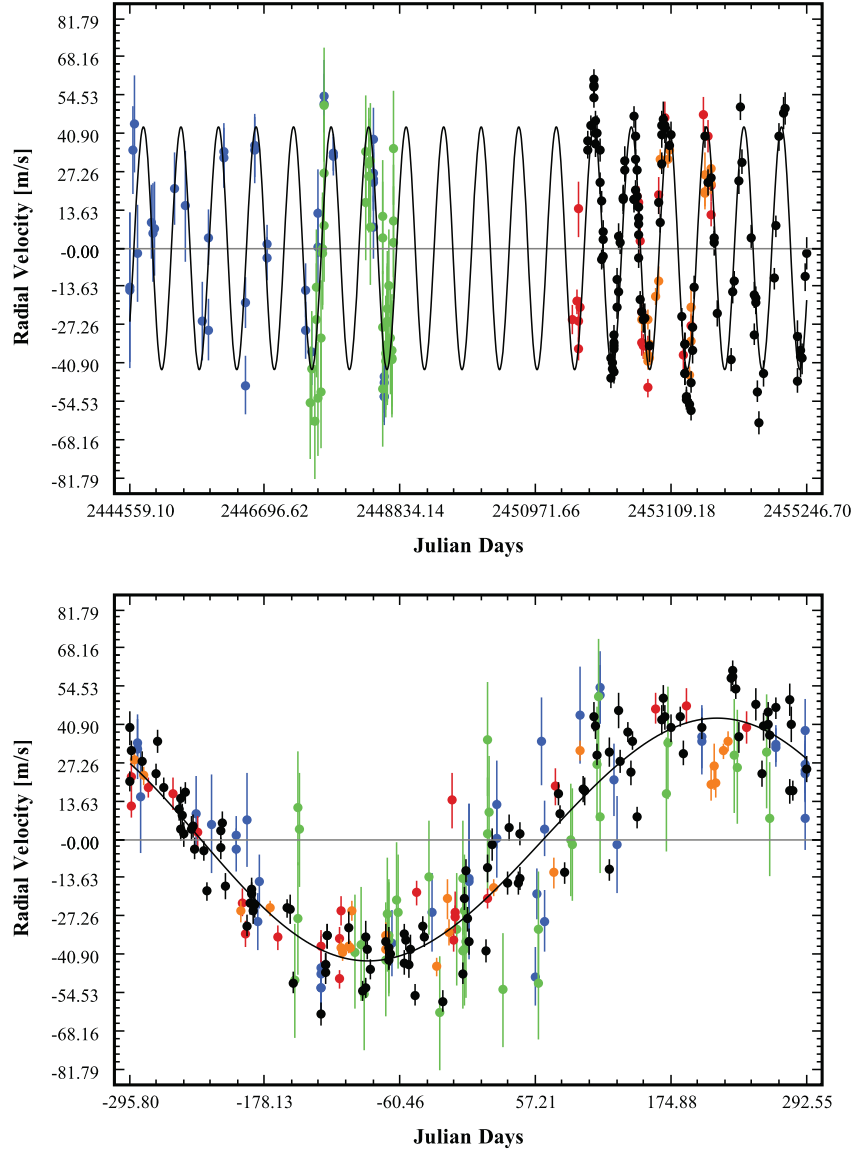}
  \caption{\label{Pollux} Radial-velocity measurements and Keplerian orbital fit for Pollux ($\beta$\,Gem). Data points plotted in blue are from the CFHT (Larson et al.\ 1993), green data points are from McDonald Observatory (Hatzes \& Cochran 1993), red and orange data points are from McDonald and Tautenburg (Hatzes et al.\ 2006), the black points from Lick (Reffert et al.\ 2006 and new measurements). The top panel shows data as a function of time, in the bottom panel all data are folded to a period of 592.6\,d.}
\end{figure}

A nearly sinusoidal radial-velocity variation has been detected for Pollux (Reffert et al.\ 2006 and Fig.~\ref{Pollux}). This variation has been stable in phase and amplitude for almost thirty years and is thus very likely due to a planet in an almost circular orbit.

\section{Statistical Inferences}

\begin{figure}
  \includegraphics[width=.75\textwidth]{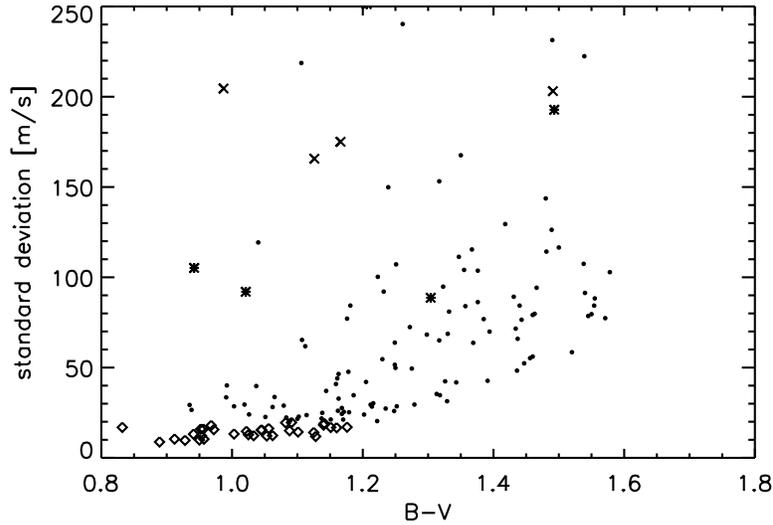}
  \caption{\label{Hekker} Standard deviation of the radial velocity of the 179 original stars from the Lick survey
survey plotted as a function of $B-V$. Most stars with $B-V < 1.2$ show
smaller variations in the radial velocity than the ones with $B-V > 1.2$.
The diamonds represent the ``stable'' stars, the asterisks represent the binaries, and
the crosses variable stars with a long-term trend. Stars with a standard
deviation larger than 250\,m/s, mainly binaries, are not shown. From Hekker et al.\ (2006).}
\end{figure}

For a statistical analysis of the survey results it is vital to separate companion detections from spurious signals due to stellar ``noise'', i.e., photospheric variations due to oscillations or rotational modulation. Figure~\ref{Hekker} shows the radial-velocity rms of our 179 original stars as a function of their color, $B-V$. It is obvious that redder stars generally show larger variations. In particular, there is a lower envelope to all data points indicative of the level of Solar-like oscillations, whose amplitude increases with color, and which are undersampled in our data. One will thus have to expect that any planet signatures will be superposed on intrinsic stellar noise, and that the amplitude of this noise will depend on the stellar color and perhaps other stellar properties.

\begin{figure}
  \includegraphics[width=.77\textwidth]{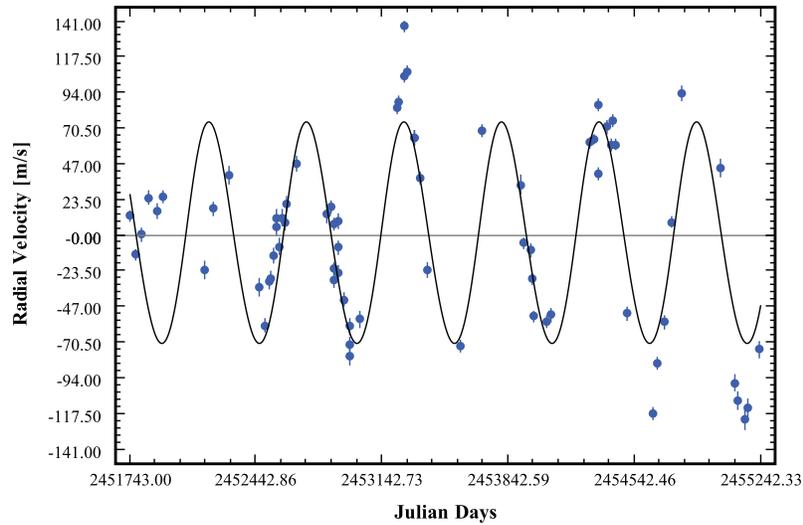}
  \caption{\label{upsper} Radial-velocity data of $\upsilon$\,Per, with a tentative Keplerian fit ($p = 541$\,d, $e = 0.039$). There are clear systematics in the residuals, which cast doubt on the companion interpretation of the radial-velocity variations.}
\end{figure}

In individual cases, it may be rather difficult to distinguish between companions and intrinsic stellar variations, as demonstrated by the example of $\upsilon$\,Per (Hekker et al.\ 2008, see also Fig.~\ref{upsper}). The radial-velocity data of this star show a significant periodicity of 541 days, but there are clear systematics in the residuals around the best Keplerian fit. It is very hard to tell whether this means that there is a companion superposed on some lower-level intrinsic RV variations, or whether these variations are entirely due to a photospheric mechanism. One has to be particularly careful with data that cover only one or two full periods, because a short sequence of random variations can easily mimic a periodic signal. One can in principle use additional information such as photometric data, analyses of line shapes, or variations of lines that are good activity indicators to distinguish between the two possibilities. However, the absence of variations in these indicators cannot conclusively prove the reality of a companion unless one has quantitative models for all alternative scenarios linking the amplitude of radial-velocity variations with those expected in the complementary data.

\begin{figure}
  \includegraphics[width=.75\textwidth]{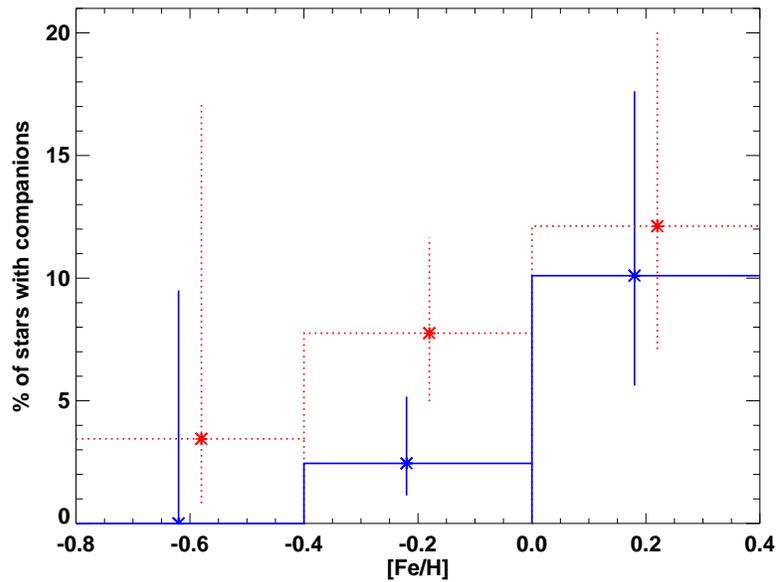}
  \caption{\label{hist} Companion frequency as a function of metallicity, [Fe/H]. The blue (lower) histogram shows secure companion detections. The red (higher) histogram includes tentative candidates. The error bars denote uncertainties due to Poisson statistics (95\% confidence intervals).}
\end{figure}

Because of these difficulties we have decided to introduce two categories of companion detections for the purpose of statistical analyses: secure ``companions'' (such as $\iota$\,Dra) and tentative ``candidates'' (such as $\upsilon$\,Per). For the definition of the first category we apply strict criteria, whereas we liberally include periodic signals with unclear origin in the second. This should ensure that the true companion frequency is bracketed from below by that of the first class, and from above by the sum of ``companions'' and ``candidates''. Figure~\ref{hist} shows these two frequencies for our giant sample as a function of metallicity. There is a clear trend of the companion frequency increasing with metallicity, quite like in main-sequence F, G, and K stars (e.g., Santos et al.\ 2004, Fischer \& Valenti 2005), and contrary to the claim that such a trend is absent in giant stars (Pasquini et al.\ 2007). This conclusion is evidently rather independent on the criteria applied to distinguish between planets and stellar noise; almost any curve between the lower and upper lines in Fig.~\ref{hist} increases significantly to the right. There is one important caveat, however: The details of the sample selection criteria might have a significant influence on correlations as the one between companion frequency and metallicity considered here. In general stellar properties such as mass, metallicity, surface gravity, evolutionary state, amplitude of Solar-type oscillations, starspot coverage, color, and absolute magnitude are all correlated with each other, and each of these parameters may influence either the frequency of companions or the detection threshold (or both); one would have to perform a multi-dimensional analysis to separate the roles of these parameters from each other. The available sample sizes are barely large enough to draw firm conclusions about planet frequency in the mass-metallicity plane for main-sequence stars and subgiants (Johnson et al.\ 2010); the inclusion of giant star samples in this type of analysis requires good statistics and careful consideration of the effects of the sample selection criteria.

\section{The $\nu$\,Oph System}

\begin{figure}
  \includegraphics[width=.77\textwidth]{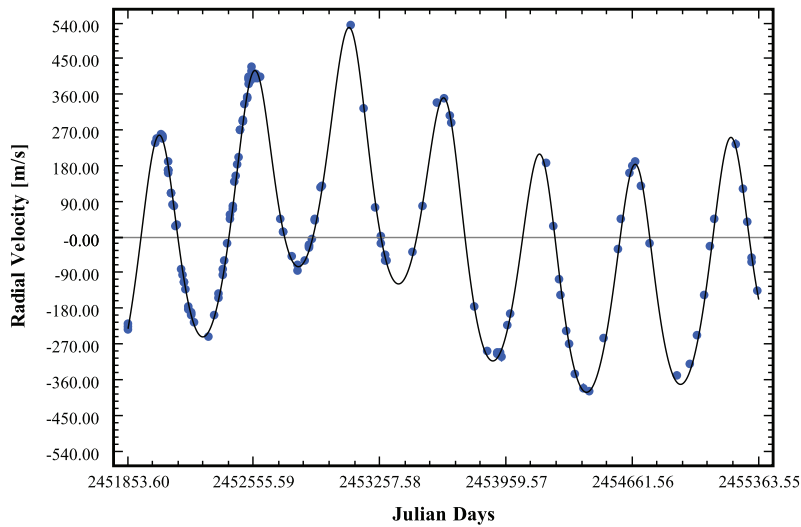}
  \caption{\label{nuoph} Radial-velocity data of $\nu$\,Oph, with double-Keplerian fit. Fit parameters are given in Tab.~\ref{nuophtab}.}
\end{figure}

One of the most interesting stars in our sample is $\nu$\,Oph. Mitchell et al.\ (2003) announced the discovery of a brown dwarf companion to this star with $m \sin i = 22$\,M$_{\rm jup}$ based on initial data from our survey. Continued monitoring quickly showed that a single Keplerian cannot fit the data satisfactorily, and the full eleven-year data set now clearly reveals the presence of a second companion with slightly larger mass in a nine-year orbit. Figure~\ref{nuoph} shows the radial-velocity data for this star along with a double-Keplerian fit, whose parameters are given in Tab.~\ref{nuophtab}.

\begin{table}[t]
\begin{tabular}{cccc}
\hline
Companion & $P$ [days] & $m \sin i$ [M$_{\rm jup}$] & e \\
\hline
$\nu$\,Oph\,b &  530 & 22.3 & 0.13 \\
$\nu$\,Oph\,c & 3169 & 24.5 & 0.18 \\
\hline
\end{tabular}
\caption{Parameters of the $\nu$\,Oph system.}
\label{nuophtab}
\end{table}

The mass of $\nu$\,Oph is about 2.7\,M$_\odot$. Brown dwarf companions appear to be much more abundant around giants in this mass regime than around Solar-type main-sequence stars, but to our knowledge this is the first case of a star orbited by two brown dwarfs. Even more remarkably, the period ratio between the two orbits is very close to 6:1.

It is not clear a priori how systems consisting of a star and a brown dwarf form.\footnote{We follow the working definition of the IAU that any substellar object with mass above 13\,M$_{\rm jup}$ should be called a brown dwarf independent of its formation mechanism.} One possibility is the formation through a binary star formation process, in which case the system could be regarded as a binary star with extremely large mass ratio. Another possibility is a ``planet-like'' formation process in a circumstellar disk, either through core accretion or through a disk instability. The $\nu$\,Oph system may provide a possibility to distinguish between these scenarios. It seems plausible that the system evolved into the 6:1 configuration by migration in a disk, analogous to the favored explanation for systems in mean motion resonances (e.g., Kley et al.\ 2004). If this interpretation is correct, the $\nu$\,Oph system would show that objects with masses $> 20$\,M$_{\rm jup}$ (i.e., brown dwarfs, which one might also call ``super-planets'') form rather frequently in the disks of young intermediate-mass stars.

\section{Conclusions}

Radial-velocity surveys of giant stars can provide important insights into planet formation processes, as they can probe intermediate-mass stars, which are difficult targets for radial-velocity measurements while they are still on the main sequence. Giant star surveys have yielded a rather large number of massive planets, indicating that the frequency of these objects is larger for stars with masses around 2\,M$_\odot$ than for Solar-mass stars. One has to be careful, however, to distinguish between radial-velocity signals induced by companions and intrinsic photospheric variability, and one has to take sample selection effects into account for statistical analyses. With these caveats in mind, there is strong evidence that the trend of higher planet frequency with increasing metallicity extends into the mass range around 2\,M$_\odot$. Individual systems can also place strong constraints on planet formation scenarios. In this respect the $\nu$\,Oph system should be of particular interest, because it may provide evidence for a formation process for brown dwarfs with masses above 20\,M$_{\rm jup}$ in a circumstellar disk.


\begin{theacknowledgments}

We thank the staff of Lick Observatory for their excellent support over many years. David Mitchell, Saskia Hekker, Simon Albrecht, Christian Schwab, and Julian St\"urmer have spent many nights on Mt.\ Hamilton to collect giant star spectra. Special thanks are due to Geoff Marcy, Paul Butler, and Debra Fischer for permission to use their equipment and software.

\end{theacknowledgments}





\end{document}